# The impacts of congestion on automobile accessibility. What happens in large European cities?

MOYA-GÓMEZ, BORJA (*)

Transport, Infrastructure and Territory Research Group (t-GIS)

Human Geography Department

Faculty of Geography and History

Universidad Complutense de Madrid (UCM)

c/ Profesor Aranguren S/N, 28040 Madrid, Spain

Phone: +34 91 394 57 51

e-mail: bmoyagomez@ucm.es

GARCÍA-PALOMARES, JUAN CARLOS

Transport, Infrastructure and Territory Research Group (t-GIS)

Human Geography Department

Faculty of Geography and History

Universidad Complutense de Madrid (UCM)

c/ Profesor Aranguren S/N, 28040 Madrid, Spain

Phone: +34 91 394 59 52

e-mail: jcgarcia@ghis.ucm.es

* Corresponding author

## ABSTRACT

Every day, a large part of the population in big cities suffers transport congestion. One effect of this is a change in the spatial distribution of accessibility, which may lead to people or businesses finding themselves temporally in areas where accessibility values are lower than those that are either desired or required. This article studies changes in automobile accessibility over the course of the day caused by congestion of the road network in eight abstractmetropolitan areas of the European Union: London, Paris, Madrid, Berlin, Barcelona, Roma, Hamburg and Milano. The study has been carried out using millions of data points on real speeds on segments of the road networks gathered throughout two years from TomTom® devices, which enables a dynamic perspective on accessibility to be incorporated. In each of the areas studied, the different impacts of congestion on automobile accessibility can be observed from differences in the distribution of opportunities and the provision of infrastructures, as well as from differences in culture and customs. Despite these differences, all cities experience two peaks with a lower value during the morning and afternoon. However, results show differences in the intensity and form of the effects of congestion on accessibility in these metropolitan areas. London, Paris and Roma are the cities where congestion has the greatest impact on automobile accessibility, while the Spanish cities are hardly affected by it.

## 1. INTRODUCTION

Congestion is a problem of the land use/transport/society system. It is usually associated with large metropolitan areas, where the land is a very scarce and very valuable resource owing to the basis of their development: the high concentration of people, activities and services, their interchanges, and the economies of scale (Rode et al., 2014). Congestion seems to be inherent to agglomeration. As a consequence, cities usually demand that the infrastructure networks, and their auxiliary venues such as parking lots, consume the minimum but necessary land, in order to allow them to function properly. A notable aspect of this trade-off in land-use distribution is the tolerance of congestion, since this may become a major obstacle for the development of cities. Some authors assert that the limits of a city or urban region are determined by congestion (Gospodini, 2002; Turok and Mykhnenko, 2007) as the benefits from the concentration of activities may not be sufficient to compensate for the congestion costs (Batty, 2008; Louf and Barthelemy, 2014).

In 2011, each automobile commuter in the major cities of the United States spent 38 hours and 19 gallons (~72 litres) of petrol as a result of congestion, a cost of $818 per traveller for the year (Schrank et al., 2012). In the biggest cities of the European Union (EU), time spent

due to congestion over the year for standard 30-minute automobile journeys in 2012 ranged from the 59 hours observed in Madrid to 97 hours in Paris and Roma (TomTom, 2013a). In economic terms, the annual cost of congestion in the EU has been estimated to be 1% of GDP (Christidis and Ibáñez Rivas, 2012). All these values are based on free flow speed, i.e. they are the upper congestion cost limit. Congestion also has other negative externalities, such as increased levels of noise, pollution and the potential for accidents and reductions in the useful life of vehicles (OECD, 1999) and the capacity of the network to cope with incidents (its resiliency). Furthermore, although congestion is associated with private transport, it could severally affect public road transport services and the social groups dependent on them as well.

Since the temporal imbalance between demand and infrastructure capacity creates congestion (Ortúzar and Willumsen, 2011), many solutions have been based on increasing network capacity, e.g. new lanes or roads. Unfortunately, this type of solution may perpetuate the problem and even exacerbate or spread it to other parts of the network and other relationships. It could also create new problems, interfere with bus and pedestrian itineraries, or damage natural ecosystems (Litman, 2014). Such solutions tend only to sustain the "vicious circle" of congestion (Handy, 1993). Furthermore, the conventional tools used to evaluate such solutions - that is, transport models - do not usually assess reactions, i.e. they omit the reaction and response times required by different members when faced with new situations (Gifford, 2003; Straatemeier, 2008).

Congestion must be *managed*, consequently it is essential to find land use/transport policies and suitable indicators that would allow changes in the way the whole system functions, and make it more sustainable by inducing changes in individual behaviour. But what are the policies that can achieve such objectives? How can their effects and results be measured? These questions are not to be answered lightly as they may produce apparently good solutions that in reality have undesired effects (Levine and Garb, 2002). A good starting point for defining and controlling these policies is to measure the quality and intensity of spatial-temporal availability on activity participations/interactions and its evolution: *Accessibility* could be a good candidate. Accessibility is defined as "the extent to which the land use/transport system enables (groups of) individuals or goods to reach activities or destinations by means of a (combination of) transport(s) mode(s)" (Geurs and van Wee, 2004, p. 128), "the ease with which activities can be reached, given a location, using a specific transport system" (Morris et al., 1979, p. 91), or the ease of interaction with a significant number of opportunities (Breheny, 1978; Bruinsma and Rietveld, 1998; Hansen, 1959).

Accessibility and congestion are linked by mobility. Accessibility, induces mobility (Giuliano, 1991; Handy, 2005; Thill and Kim, 2005) and its possible excess (Salomon and Mokhtarian, 1998), and mobility could create congestion. As a consequence, congestion temporally reduces accessibility values (Mondschein et al., 2011), since the activities' location are usually immobile and their reachability only depends on the transport network performance. It is worth mentioning that "accessibility and mobility are often used together in transportation plans but without clear distinction" (Handy, 2002, p. 3), but "they convey fundamentally different concepts" (Ross, 2000, p. 13) . The aquarium allegory (Neutens et al., 2011) might clearly show the difference between both: accessibility shapes the spatial-temporal prism (or aquarium) according to opportunities/activities spatial distribution, schedules and their minimum participation time, the time and individual-based transport network performance, and so on: it is the visualization of the potential travel and activity participation (Miller and Bridwell, 2009); i.e. accessibility is a potential value, it does not point any chosen choice but evaluates the sum of possible choices in one or more means of transport. While mobility measures the characteristics and quantity of every movement in this prism (the sequence of chosen choices), to reach an available location from another one for a specific starting time, and the sum of all of them at same time could modify transport network performance, i.e. congestion.

Previous studies have shown the heterogeneous impact of congestion on the spatial distribution of accessibility (Bertolini et al., 2005; Lei and Church, 2010; Vandenbulcke et al., 2009). It could trigger some sort of reaction in members of the system (Sweet, 2011): from changes in the travel routes or schedules -short term decision- to land use relocation to more resilient spaces -long term decision (de Abreu e Silva and Goulias, 2009; Levinson and Kumar, 1994; Sweet, 2014).

This paper only focus on a part of the accessibility measures: the effects of recurring congestion, i.e. that is produced despite the absence of incidents (Stopher, 2004), on the automobile accessibility values of eight large metropolitan areas in the EU. It makes use of network speed data taken from TomTom devices. With this information it is possible to work with the same definitions for all the study areas and obtain more. It is also possible to know with great accuracy the temporal accessibility variations with respect to territory, population and inequalities.

The paper is structured as follows: Section 2 shows the used methodology on this paper. Section 3 explains the characteristics of the study areas and the details of the data of the network used. The results obtained are given in Section 4. The last section discusses conclusions and possible future research.

## 2. METHODOLOGY

Congestion is a dynamic phenomenon that requires dynamic data and appropriate methodologies to study it properly (Ben-Akiva, 1985). Previous studies on automobile accessibility and its variation have been limited to using static scenarios (see (Tilahun et al., n.d.; Vandenbulcke et al., 2009)). This methodology is suitable for studies in which the all properties of the network links can be considered constant throughout the duration of any trip. It tends not to include adequately the consequences of congestion or its temporal dimension: travel time depends not only on the origin, destination, transport mode and route chosen, but on the moment each link of the network is used and its temporal impedance. New data sources and adapting traditional methodologies to dynamic reality, it is possible to overcome the conceptual limitations of using static scenarios.

On this paper, we use a potential accessibility (Hansen, 1959) zone-based indicator to measure the effects of congestion on territorial automobile accessibility. We used a negative exponential function, its main point of interest is the transformation of all opportunities into MPUs in cases where the trip incurs no costs, and it does not require any mathematical artifice. This way, we also avoided the effects of the self-potential problem (Frost and Spence, 1995). The temporal congestion component was only incorporated in the indicator through dynamic estimation of the shortest travel time route, for each origin-destination relationship and different instances of departure. Opportunities remained constant.. Equation A shows the definition of accessibility and cost estimation used, which follows the Weibull's axiomatic approach of accessibility measurement (Weibull, 1976).

$$A_i^t = \sum_{j \in N} D_j \cdot e^{\beta \cdot c_{ij}^t} ; \forall\, i \in N, t \in T \quad (A)$$

subject to:

$$c_{ij}^t = \sum_{m \in M} \sum_{e \in E} \alpha_{eij}^{tm} \cdot c_e^m ; \forall\, ij \in G, t \in T$$

Where:

$A_i^t$ is the potential accessibility value of origin *i*, beginning at instant *t*.

$D_j$ is the opportunities of destination *j*

$e^{-\beta \cdot c_{ij}^t}$ is the impedance-decay function.

β is the used parameter. In our case[1], we used β= -0.065.

$c_{ij}^t$ is the impedance experienced when travelling from origin *i* to destination *j* by the shortest route, beginning at instant *t*. On this paper, the impedance is the travel time [min].

$\alpha_{eij}^{tm}$ is the binary variable that indicates whether network link *e* is used for the trip between origin *i* and destination *j* which has begun at instant *t*, starting its use at instant *m*,

$c_e^m$ is the expected impedance of network link *e*, use of which begins at instant *m*. On this paper, the expected time is the travel time [min]

$N$ represents all the zones included in the calculation area.

$G$ is the set of origin-destination relationships, including relation with itself (origin i = destination j)

$T$ is the set of instants of started trips.

$M$ is the all possible instants within the study.

We assume that the network within every scenario has *First In First Out* (FIFO) properties, i.e. there is no overtaking. The estimation of routes in dynamic FIFO networks can be resolved with very slight modifications to the algorithms used in static networks (Chabini, 1998; Dean, 2004).. In spite of the FIFO consideration is not satisfied at micro level (vehicle to vehicle), it is considered adequate at meso and macro levels (flows).

By applying Equation A to different departure instances and all origin-destination (O/D) relationships, we obtain the necessary data for estimating temporal accessibility. In this paper, we calculated 96 accessibility values for a day, i.e. one every 15 minutes. This makes it possible to estimate the daily variations in accessibility distribution in our study cases.

---

[1] The parameter was estimated with the OD matrix and travel times in the cases of Madrid and Barcelona. This value is constant for all the cities in order to compare them all under the same conditions of calculation.

Finally, in order to understand each metropolitan area as a unit and carry out comparisons between them, we calculated the global accessibility profile, as expressed in the weighted average (Equation B):

$$A_{global}^{t} = \frac{\sum_{\forall i \in N} A_i^t \cdot O_i}{\sum_{\forall i \in N} O_i}; \forall\, t \in T \tag{B}$$

Where:
$A_{global}^{t}$ is the global weighted accessibility value of the study area, when trips start at instant *t*.
$A_i^t$ is the accessibility value of zone *i* when trips start at instant *t*.
$O_i$ is the weight or potential of origin *i* (in our case: inhabitants).
*N* is all the study zones.
*T* is all the instants of trips started.

After global analysis, we analysed the spatial distribution of the effects of congestion at local level in each metropolitan area. This was carried out by mapping, according to the zones of origin, the results of the maximum accessibility value (in free flow), the relationship between the accessibility value in free flow and accessibility values at morning and afternoon congestion peaks, and the moment of least accessibility in each zone were also identified. Finally, each zone is classified into the most similar general profile identified for each of the eight study areas. To do the assignation, we used the "kml" R-project[2] package for longitudinal data classification procedures.

All the processes were carried out using ESRI® ArcGIS 10.1 for generating networks[3], calculating different impedance matrices with their algorithm in the heuristic version for dynamic FIFO networks, and for including turn restrictions and directions in the network. We also used different ad-hoc Python scripts for mass processing the data on routes, and the

---

[2] https://cran.r-project.org/web/packages/kml/index.html3 The free tool for ArcGIS StreetDataProcessingTools[3] was used to create correctly the Network Datasets with TomTom® data. Available for download at:
http://www.arcgis.com/home/item.html?id=755f96fcde454ece8f790fecb3e031c7
[3] The free tool for ArcGIS StreetDataProcessingTools[3] was used to create correctly the Network Datasets with TomTom® data. Available for download at: http://www.arcgis.com/home/item.html?id=755f96fcde454ece8f790fecb3e031c7

R-project for different statistical studies.

## 3. THE STUDY AREAS. EIGHT EUROPEAN CITIES

On this paper, we studied the effects of congestion on automobile accessibility in eight of the most populated metropolitan areas in the EU: London (the United Kingdom), Paris (France), Madrid (Spain), Berlin (Germany), Barcelona (Spain), Roma (Italy), Hamburg (Germany) and Milano (Italy). Although each area shows different characteristics with respect to infrastructure, distribution of opportunities, legislation, and the customs and habits of its residents, they are all typified by the existence of an unmistakable core city that brings together the whole metropolitan space. This does not mean that no other important cities are found within the metropolitan areas under study.

In this section and the next one, we will use "accessibility" to refer to "automobile accessibility". Any reference to trip or access should be interpreted as travelled/observed by car mode.

### 3.1. Study area delimitation

On this paper, each study area is taken to be all the LAU2 (Eurostat, 2011), also known as municipalities, that belong to the Functional Urban Area (FUA, (ESPON, 2014)) of the main city or of any other FUAs completely surrounded by the main one and have more than 50% of their area within the density isoline of 500 inhabitants/km$^2$ of the main city. This isoline was generated with the *density kernel* tool[4], using the 1 km$^2$ EEA reference grid (European Environment Agency, 2007; Peifer, 2011) with Eurostat population data from 2006 (Eurostat, 2006). In the case of London, with its particular administrative system, it was decided to define the study areas in accordance with LAU1. Table 1 shows general information on each of the study areas, while Figure 1[5] shows population distribution in 2006 (Eurostat, 2006).

| City[6] | Country | Code city | Downtown | Num. LAU2 | Pop. of Study Area (inh) | Total Area of Study Area (km$^2$) | Car share |
|---|---|---|---|---|---|---|---|
| London | United Kingdom | LON | Charing Cross | 70* | 11,719,462 | 6,540 | ~40% |
| Paris | France | PAR | Notre Dame | 500 | 10,473,157 | 3,508 | ~35% |
| Madrid | Spain | MAD | Puerta del Sol | 39 | 5,502,282 | 2,312 | ~30% |
| Berlin | Germany | BER | Alexanderplatz | 27 | 3,905,813 | 1,840 | ~35% |

---

4 Density kernel is a tool of ArcToolBox of ArcGIS; the search radius used to estimate the values was 10.000 metres.
5 The projection of all maps in this paper is the ETRS89 - Lambert Azimuthal Equal Area (LAEA) Europe (EPSG: 3025). All of them are on the same original scale (1:500,000 in DIN A4). They are also available in electronic supplementary material.
*6* The cities are sorted by the most populated LUZ (ESPON, 2014) to less one.

| | | | | | | | |
|---|---|---|---|---|---|---|---|
| Barcelona | Spain | BCN | Plaça de Catalunya | 88 | 4,277,836 | 1,420 | ~40% |
| Roma | Italy | ROM | Piazza del Campidoglio | 21 | 3,046,642 | 1,992 | ~60% |
| Hamburg | Germany | HAM | Rathuis | 41 | 2,294,105 | 1,504 | ND |
| Milano | Italy | MIL | Piazza del Duomo | 339 | 4,883,774 | 2,824 | ND |
| *(\*) LAU1* | | | | | | | |

Table 1. General information on the study areas. Car share data source (UN HABITAT, 2013)

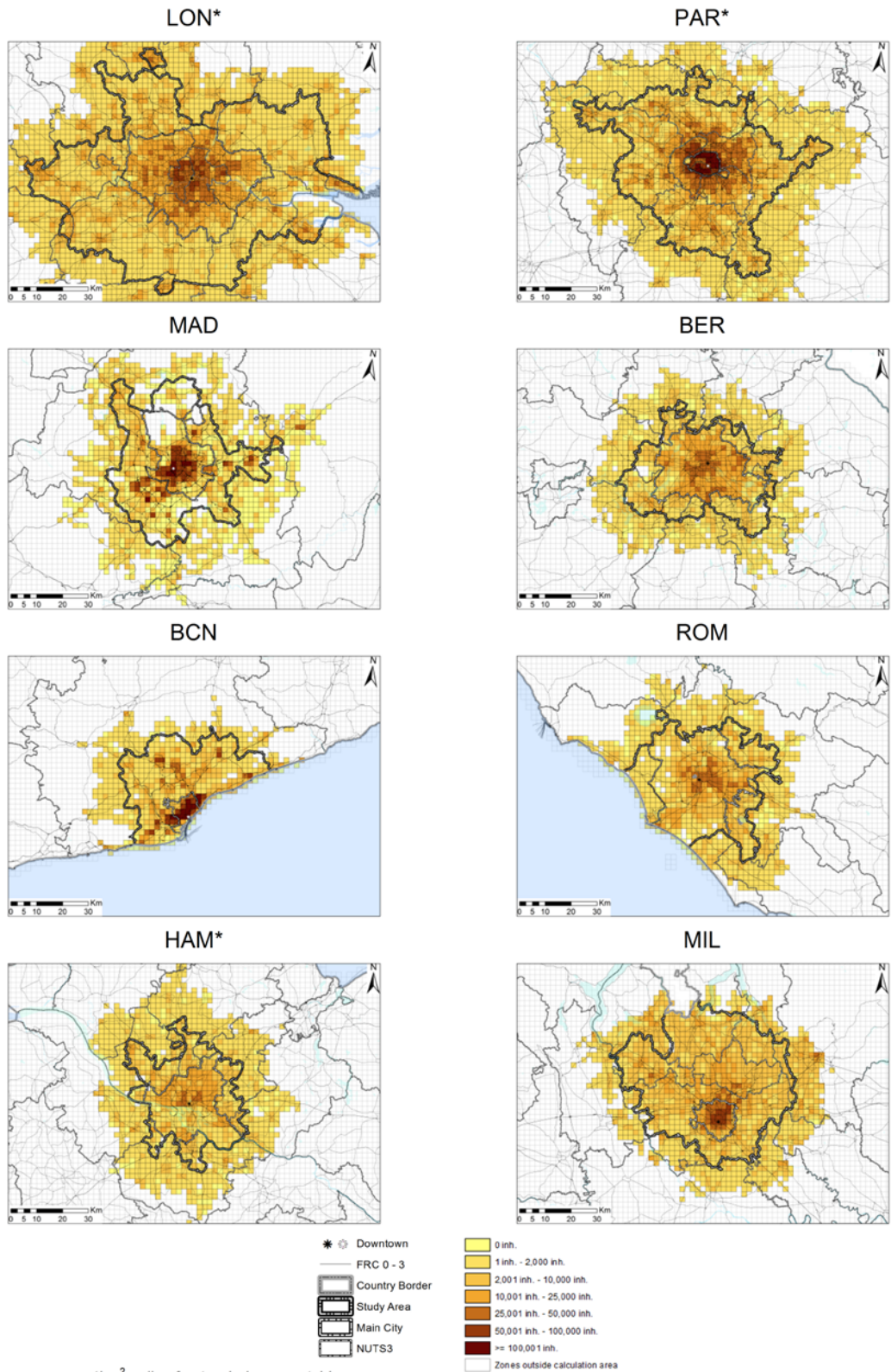

Figure 1. Spatial distribution of population of study areas and area of calculation (2x2km cells)

The maps show the different urban morphologies. There is a marked tendency in Madrid and Barcelona towards a concentration of inhabitants in the main city, while Hamburg, Berlin and Roma have a more sprawled distribution. Paris, Milano and London are intermediate cases, ranging from a relatively compact city in the case of Paris to a more diffuse and extensive one in the case of London. In all cases, there are major towns close to borders that are well connected with the central city. These usually have similar distribution patterns to the main city. Figure 2 summarizes these comments and shows how total population and net population density varies from each downtown.

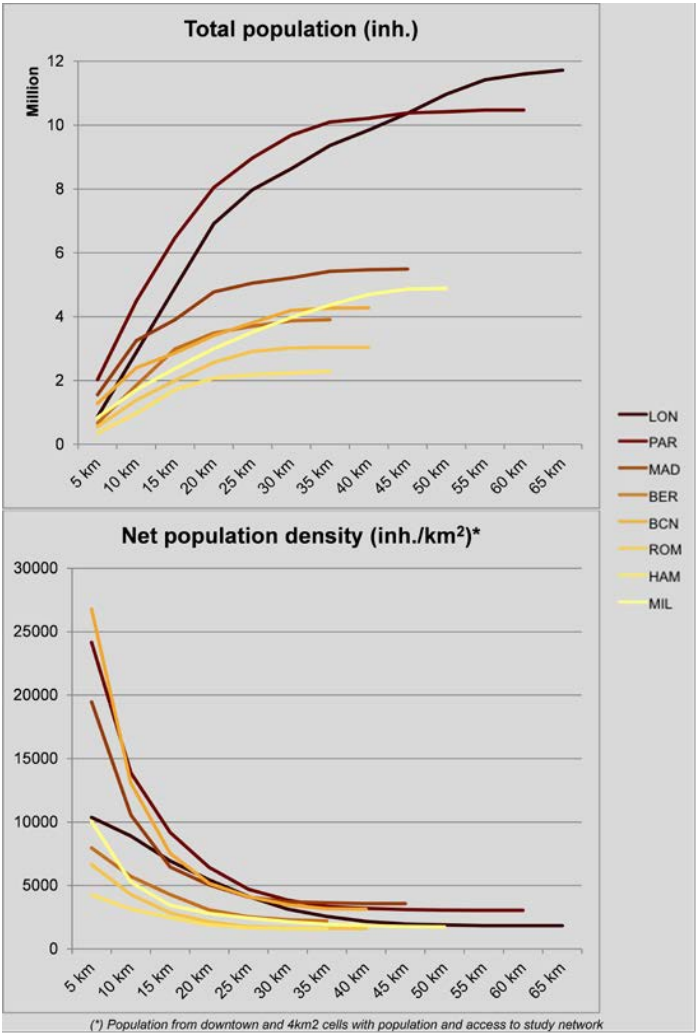

Figure 2. Accumulate population and net population density from each downtown

## 3.2. The Origin/Destination zones data ($O_i$ & $D_j$)

We divided each study area into a regular grid of 2x2 km (4km$^2$) cells based on the 1 km$^2$ EEA reference grid; they became the Origin/Destination zones. By using regular grid data from the same source it was possible to overcome some aspects of the Modifiable Areal Unity Problem (MAUP) (Kwan and Weber, 2008) and carry out comparisons of the study areas without previous treatments being required. On the other hand, we used some outside study area cells –which can be reached from study area at most at 15minutes at midnight, to reduce border effects.

Values of opportunities at destinations are constant in time in order to only focus our study on congestion effects (Geurs and Van Eck, 2003). The population has been used as a representative static value of the opportunities at each destination throughout the day. These values were also used for weighting global accessibility. Table 2 shows the total cells considered and the routes generated at each moment analysed in time for the metropolitan areas.

| City | Total calculation cells | Total routes for each starting instant |
|---|---|---|
| London | 2,837 | 8,048,569 |
| Paris | 2,018 | 4,072,324 |
| Madrid | 1,236 | 1,527,696 |
| Berlin | 1,113 | 1,238,769 |
| Barcelona | 604 | 364,816 |
| Roma | 973 | 946,729 |
| Hamburg | 1,119 | 1,252,161 |
| Milano | 1,257 | 1,580,049 |

Table 2. Summary of O/D data per city

We have to recognize that using population as opportunities and weighting values, especially in urban studies, could not get whole system complexity and the derived behaviour. We should also take into account other accessibility values, such as jobs, scholar, medical or shopping. However, population accessibility on metropolitan areas could be useful to urban freight distribution – their vehicles represent 8-15% of total traffic flow in European metropolitan areas (European Commission, 2012), emergency services studies, or face-to-face interactions - some authors even argue that social interactions are the base of successful cities (see Sim et al., 2015), among other topics.. Moreover, population data is the unique data at 1km$^2$ grid level for all cities from the very same data source, whereby it avoids some unacceptable biases in city comparison.

## 3.3. TomTom®'s Speed Profiles. The data of network performance ($c_e^m$)

On this paper, we use networks formed by links in categories 0 to 6 of the Functional Road Classification (FRC)[7] for TomTom® road networks (March 2013 version, data from 2011 and 2012). They incorporate all roads and the main urban network, and almost all the available information of TomTom®'s Speed Profiles data.

The Historical Speed Profile is made up of the speed values observed every 5 minutes (TomTom, 2013b), as a percentage of the speed at each moment with reference to the observed free flow speed (Figure 3). Each link was assigned one of the 98 predetermined profiles for each weekday. Each profile might have different values to 100% between 4:30 and 21:20 h (16 hours and 50 minutes). This data structure saves on memory and is devised to be suitably treated in an ESRI® ArcGIS environment (ESRI, 2014). On this paper, the information used relates to Wednesdays.

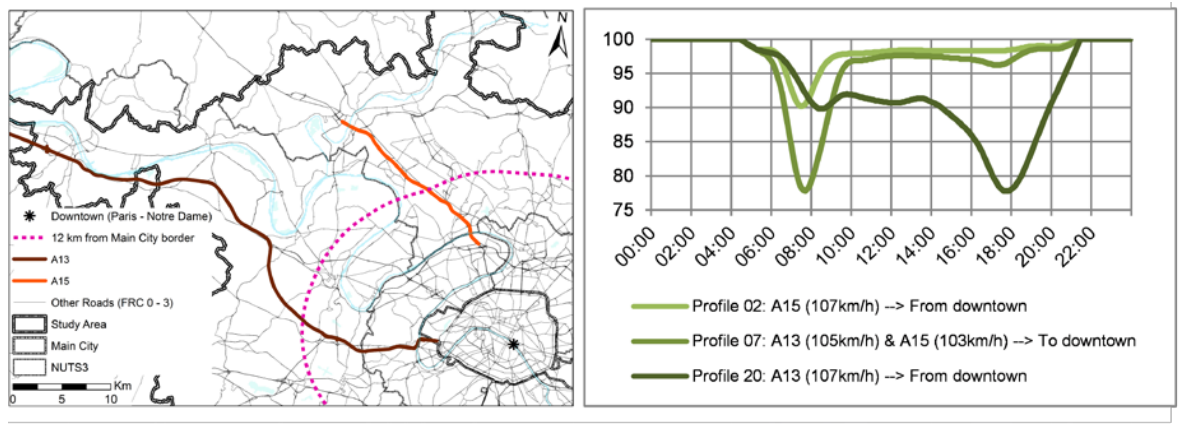

Figure 3. Some TomTom®'s Speed Profiles for Wednesdays in A13 and A15 at 12km from Main City boarder (Paris)

## 4. WHAT EFFECT DOES CONGESTION HAVE ON THE AUTOMOBILE ACCESSIBILITY OF EUROPEAN CITIES?

This section shows the results of changes caused by congestion on the accessibility value in each of the study areas. These results show the effects of congestion on the territory over the course of the day of study (Wednesday). The first part is a global examination of the results for each study area and the second section studies the spatial distributions of the impacts of congestion on each O/D zone. The results base should in no way be understood as a recommendation for approaching congestion management policies strictly from the observed situation in free flow conditions: it is only the upper value, but it should not be the theoretical/practical optimal one. Higher resolution maps of maps shown in each section are also available in *electronic supplementary material.*

---

[7] TomTom®'s FRC Definitions. FRC 0: Motorway, Freeway, or Other Major Road; FRC 1: a Major Road Less Important than a Motorway; FRC 2: Other Major Road; FRC 3: Secondary Road; FRC 4: Local Connecting Road; FRC 5: Local Road of High Importance; FRC 6: Local Road; FRC 7: Local Road of Minor Importance; FRC 8: Other Roads.

### 4.1. Results at global level

In this section, each study area is considered as a single entity, with the aim of doing a comparison with few values (benchmarking study). Table 3 and Figure 5 show the results of value aggregation according to Equation B. Only the temporal component of the effects of congestion on each study area is introduced in this section.

| City | Max. Glob. Access. (MPUs) | Aver. Globl. Access. (MPUs) | Med. Global. Access. (MPUs) | % Aver. Globl. Access. / Max. Glob. Access. | % Med. Global. Access / Max. Glob. Access. | Morning peak | Afternoon peak |
|---|---|---|---|---|---|---|---|
| London | 1,673,785.74 | 1,438,593.62 | 1,355,431.26 | 85.95 | 80.98 | 08:00 | 17:00 |
| Paris | 2,405,382.98 | 2,101,942.51 | 1,994.151.89 | 87.38 | 82.90 | 08:00 | 17:15 |
| Madrid | 1,968,829.47 | 1,873,957.40 | 1,838,078.41 | 95.18 | 93.36 | 08:15 | 17:15 |
| Berlin | 1,094,836.39 | 986,924.79 | 963,966,78 | 90.14 | 88.05 | 08:00 | 16:45 |
| Barcelona | 1,449,408.91 | 1,362,495.35 | 1,330,630.50 | 94.00 | 91.81 | 08:15 | 17:15 |
| Roma | 922,464.54 | 811,385.60 | 763,851.27 | 87.96 | 82.81 | 08:00 | 17:15 |
| Hamburg | 753,079.52 | 679,549.70 | 658,760.67 | 90.24 | 87.48 | 08:00 | 17:00 |
| Milano | 1,272,149.57 | 1,139,023.63 | 1,091,022.37 | 89.54 | 85.76 | 08:00 | 17:15 |

Table 3. Main global accessibility results

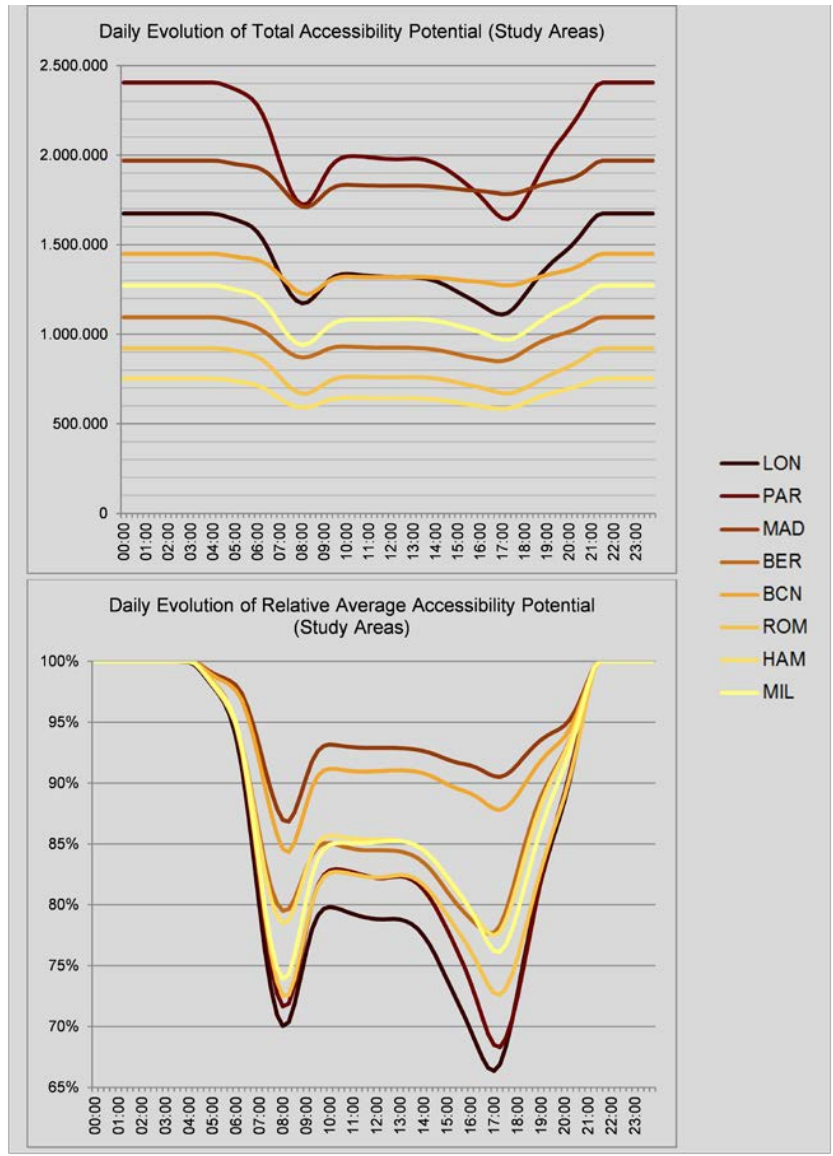

Figure 4. Daily evolution of average accessibility potential (in MPUs and %)

Global accessibility in free flow is conditioned in the eight study areas by urban morphology and network characteristics. The greatest accessibility is seen in Paris and Madrid, ahead of London, despite this being the area of greatest population. The same situation can be observed between Milano and Barcelona, with greater accessibility in free flow in Barcelona, despite its lower amount of gross opportunities.

The effects of congestion are especially marked in the two big European metropolitan areas: London and Paris (Table 3 and Figure 4). If their situation is compared to those of the Spanish metropolitan areas (the least affected by congestion) it can be seen that Barcelona, with a third of the population of London and a total area almost 5 times smaller, and London have almost the same accessibility value during the middle of the day, to the extent that Barcelona may reach an even greater number of opportunities than London during the times of greatest congestion. As a result of congestion in Paris, Madrid becomes the study area with the greatest weighted global accessibility during the afternoon congestion period. In the cases of Milano-Berlin and Roma-Hamburg, the Italian cities show greater weighted global accessibility values in free flow but their greater congestion, particularly during the morning peak, considerably reduces the differences with the German cities.

Figure 4 also shows the results as percentages with respect to the maximum value obtained and making it easier to observe the effects of congestion on accessibility. It confirms that the morning peak is more abrupt than the afternoon one, since the former takes place over 4 hours while the latter lasts for 7.5 hours (convex segment of Figure 4). What is noteworthy, however, is that there is a stationary period between the peaks, with values between 80% and 93% with respect to maximum accessibility. These values are good candidates to become accessibility reference values. During the congestion peaks, accessibility may drop by about 20-30% in the mornings and 25-35% in the afternoon. The profiles of the Spanish study areas are much gentler than the others, with one marked congestion peak in the morning, but a much smaller one in the afternoon. The two German metropolitan areas show an intermediate situation, with some profiles with almost identical minimums of accessibility and greater intensity of congestion in the afternoon. Milano shows a loss during the middle of the day similar to the German cities, but its congestion during the morning peak is much greater. Paris and Roma show a very similar accessibility profile, except that Paris loses almost 5% more accessibility than Roma during the afternoon peak. As already mentioned, London shows the worse situation. Its accessibility is reduced up to 30% during the morning peak (08:00 hours), during the middle of the day it is always below 80% of its accessibility in free flow and it shows a marked and extensive congestion peak in the afternoon (with losses of almost 35% at 17:00 hours).

### 4.2. Results at O/D zone level

This section introduces the spatial component of the impact of congestion on each study area (Table 4 and Figures 5a, 5b and 6). The situation in free flow, which would relate to traditional accessibility studies (first row, Figures 5a and 5b), shows an almost perfect concentric distribution in all the metropolitan areas analysed. Only Barcelona and Roma show some clusters of their highest accessibility values in the peripheral zones, close to main roads accessing the downtown or ring roads. As this concentric pattern is repeated in all the study areas, the differences between them are found in the rate of reduction in accessibility from the centre to the periphery. Paris, Barcelona and Milano show the most abrupt drops of all the study areas. In contrast, Roma and Hamburg have a less marked decrease, with a greater number of zones concentrating their high accessibility values in free flow.

| City | Accessibility of Max. value cell (MPUs)* | Cells with more acc. than 80% max value cell (%) | Cells worst case mornings (%) | Cells worst case mornings (%) |
|---|---|---|---|---|
| London | 2,466,611.01 | 08.96 | 71.39 | 28.61 |
| Paris | 3,619,370.42 | 06.62 | 75.00 | 25.00 |
| Madrid | 2,560,792.77 | 15.71 | 89.39 | 10.61 |
| Berlin | 1,493,730.03 | 14.16 | 67.70 | 32.30 |
| Barcelona | 1,879,698.20 | 12.50 | 91.57 | 08.43 |
| Roma | 1,231,617.81 | 20.95 | 69.09 | 30.91 |
| Hamburg | 987,158.45 | 18.92 | 61.89 | 38.11 |
| Milano | 1,847,093.91 | 10.87 | 87.41 | 12.59 |

Table 4. Accessibility behavior and values during free flow time and peaks. (* Use this value to read row 1 in Figures 5a and 5b)

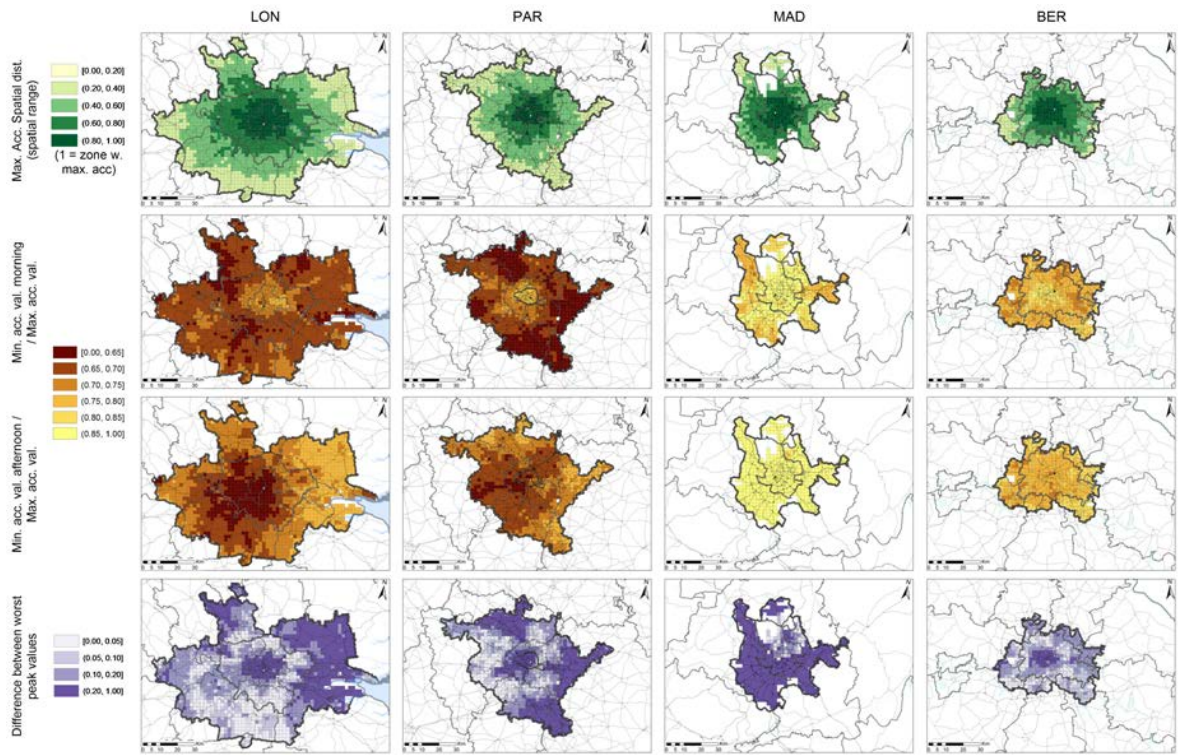

Figure 5a. Accessibility during Wednesdays in London, Paris, Madrid and Berlin

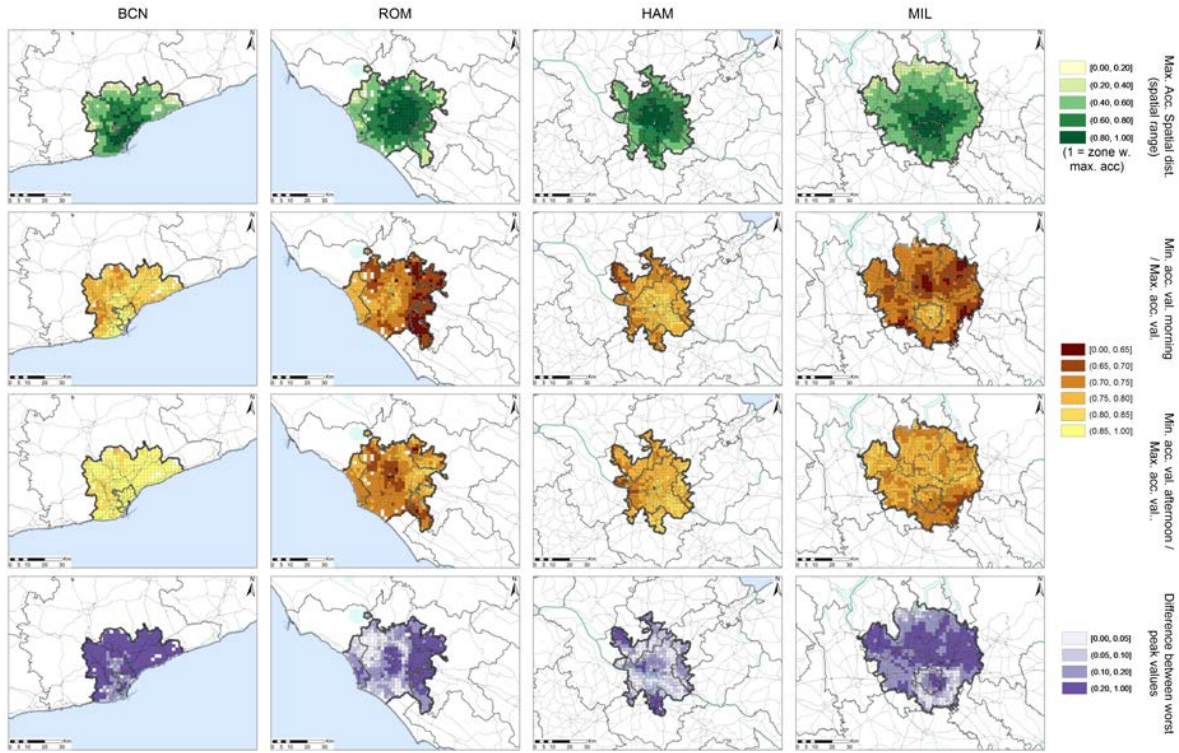

Figure 5b. Accessibility during Wednesdays in Barcelona, Roma, Hamburg and Milano

Analysis of the effects of congestion establishes a clear differentiation between peripheral zones, with worse accessibility values during the morning peak due to the flow towards the CBD and other points with a high concentration of activities for work or study purposes, and those centres. The centres and adjacent zones are mainly affected by congestion during the afternoon peak (return home). In Madrid, Barcelona and Hamburg, the centres are affected during both peaks and, although the moment of least accessibility also takes place in the afternoon (see rows 1 and 3 in Figure 6); the normalized relative difference[8] with respect to the lowest morning value is very small (see forth row in Figures 5a and 5b), Notice that a large part of German cities' zones have a small differences between worst values. This behaviour is also observed in areas surrounding the centres of the other study areas. In general, the effects of congestion spread in waves from the outside to the inside of the study

---

8 We used following equation: $\%\text{value} = \frac{\text{best.peak} - \text{worst.peak}}{1 - \text{worst.peak}}$

areas during the morning congestion period. In the afternoon, however, it is only in London, Roma, Madrid and Barcelona where the effects are seen in the opposite direction. Both London and Roma have more than one centre generating these waves (see rows 1 and 3 of Figure 6) in the afternoon, which may indicate the presence of important secondary urban centres.

Figure 6 (first and third row) shows the hours of poorest accessibility in each of the zones. There is a marked difference between the zones affected by morning congestion (fundamentally the periphery) and the zones with the greatest loss of accessibility in the afternoons (urban centres and areas of activity). If the metropolitan areas are compared, cities like London or Paris have a considerably greater number of zones that are more affected by afternoon congestion, with a slightly more delayed peak in Paris (17:15-17:30) than in London (17:00). In the Mediterranean metropolitan areas (especially the Spanish ones), however, the afternoon peak is clearly limited to urban centres and certain specific zones or corridors of activity.

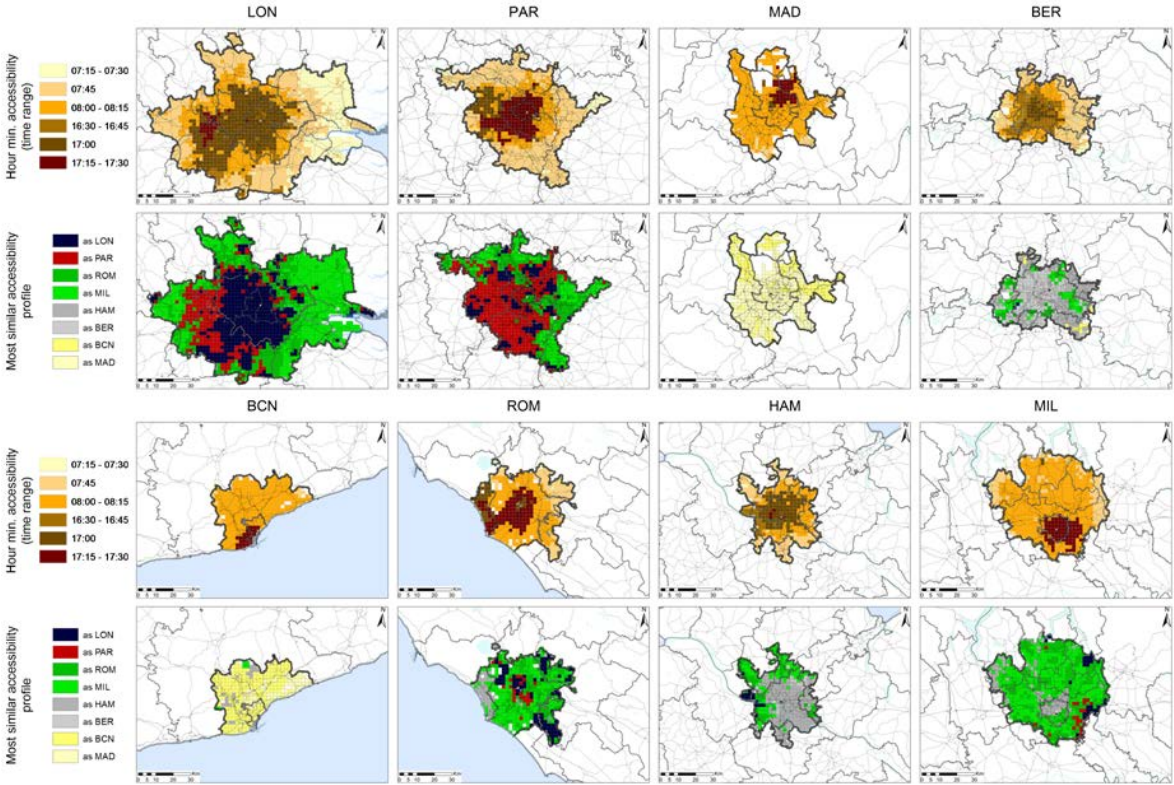

Figure 6. The worst hour and classification of each zone according the similarity of their relative profiles to the global cities relative profiles (Figure 4)

Comparative studies between zones in each of the study areas are of great interest because differences between zones might trigger some decision processes, such as reschedule and/or relocate activities. A good example of this is seen in the zones of high population concentration around the two Paris corridors: the A-13 (the western corridor from the centre) and the A-15 (the north-western corridor from the centre)[9]. The zones along the A-13 corridor show two peaks of greater impact of congestion on accessibility; nevertheless, these values are lower than the losses in the zones situated around the A-15 corridor. The same results are seen in the more peripheral zones of each study area. Similarly, "islands of zones with poor congestion resilience" can be observed, in other words, groupings of zones with losses in accessibility greater than those seen in adjacent zones.

Finally, the profiles generated for global accessibilities are used to determine which study area behaviour the profile of each zone most resembles (second and fourth line of Figure 6). Whereas the Madrid and Barcelona profiles are almost completely assigned to the zones of the study areas, the Italian profiles can be assigned to zones in 7 study areas, including the peripheries of Paris and London. The behaviour of the German areas is present only in the German areas and Italian peripheries. Finally, except for some areas of central Roma, the profiles with greatest impact (London and Paris) are only seen in their respective cities. This information is a very interesting complement to the previous ones, since it straightforward shows the time analysis for every Origin/Destination zone by using a very few known and labelled profiles, eight profiles instead more than 11,000 profiles. .

## 5. CONCLUSIONS

On this paper, the observed effects of congestion on territorial automobile accessibility and population have been mapped and analysed for eight of the most populous cities in the European Union, using data from the years 2011 and 2012. To carry this out, the definition of potential accessibility has been used to obtain a daily accessibility value every 15 minutes for each O/D zone in order to generate the respective accessibility profiles.. A dynamic perspective of accessibility has been used, in which each value obtained is not only affected by the moment of departure of all the trips that connect it to the rest of the zones, but also the moment at which each link required is used and the speed with which it can be traversed (depending on whether or not there is congestion). Although this type of study requires greater computational effort for a current single computer, both for carrying out calculations for departure times and repeating each operation throughout the study period, the results obtained from it give a clearer and more realistic idea of the impacts of congestion on accessibility, including temporal evolution and time characteristics which are unrevealed in static studies.

---

9 See Figure 3

In metropolitan areas, congestion completely distorts the spatial distribution of accessibility, which has traditionally been analysed with free flow or legal maximum speeds. The results obtained show how, depending on the study area, may undergo losses of accessibility greater than 35% with respect to the free flow situation. Furthermore, the effects of congestion have very specific spatial distribution patterns, which may result in zones with great losses in accessibility appearing next to areas close by that undergo fewer losses. In other cases, adjacent zones have different losses depending on whether it is the morning peak or the afternoon peak. Congestion may also lead to greater differences in accessibility distribution between zones with a better or worse starting situation (peripheral vs downtown)..

As already shown, study of the spatial distribution of daily automobile accessibility and the losses resulting from congestion within each study area is of great interest. Combining policies on the planning of transport systems, also including public transport, and land use is essential for mitigating the effects of congestion on accessibility and making metropolitan spaces more resilient to it. In this respect, it should be noted that the zones of each city usually behave as a whole (dominating own profiles), although there may be important differences in the distribution of opportunities and/or infrastructures performance. Consequently, it seems that cultural aspects and labour legislation also have a crucial role to play in congestion management policies and their effects on automobile accessibility.

Study of the variations in daily automobile accessibility can update these combined policies on whole transport system and land use and introduce the temporal component. This way, it is possible to locate activities in the places that are most suited to their needs for accessibility according to the different time bands in which the activities are carried out. Certain activities in places with high accessibility values in free flow can suffer unacceptable losses when there is congestion. A good example is that of access to emergency medical care that is usually by road. The location of ambulance stations must guarantee a minimum of resiliency in situations of congestion to protect the quality of the service in all time bands. Other activities could guarantee their accessibility by relocations and/or transport policies, such as time-based improvements on public transport dotation or congestion fares.

We have to recognize that the aim of this paper is only the "tip of the iceberg" on temporal studies and the use of new data in accessibility. We should use more specific data to understand better how accessibility shapes the human activities system. It could be: opportunities time variation, e.g. opening times or real time-based people distribution (Chen et al., 2011), others accessibilities values, e.g. jobs, others transport modes, e.g. public transport (Owen and Levinson, 2014), and individual abilities or characteristics among others. As a result, whole society could identify better current problems, define a reasonable target and find adequate solutions – it also means minimizing possible side effects (see (Bonsall and Kelly, 2005; Levine and Garb, 2002)) -. Moreover, next studies should compare our results to polycentric metropolitan areas as well.

Finally, it is noteworthy the dynamism of congestion is a difficult problem to explain with static maps and graphs. This paper has explored representation of the results of most significance for an understanding of the causes and consequences of this phenomenon. Even so, these representations may result in a great loss of information of interest. This is not a trivial matter, since the unequivocal simplicity and meaning of the measurements, formally known as soundness and plainness, must also be guaranteed to convert the results into useful tools for policy makers (Silva, 2013).

## 6. ACKNOWLEDGEMENTS

The authors would like to thank the Spanish Ministry of Economy and Competitiveness for funding this research as part of the SPILLTRANS, (TRA2011-27095) and DynAccess (TRA2015-65283-R) projects. We are also very grateful to the peer reviewers. Their comments were very welcome and they have improved the clearness and utility of this paper.